# Revealing the Origin and Nature of the Buried Metal-Substrate Interface Layer in Ta/Sapphire Superconducting Films

Aswin kumar Anbalagan [1, *], Rebecca Cummings [2], Chenyu Zhou [3], Junsik Mun [2,3], Vesna Stanic [4], Jean Jordan-Sweet [4], Juntao Yao [2,5], Kim Kisslinger [3], Conan Weiland [6], Dmytro Nykypanchuk [3], Steven L. Hulbert [1], Qiang Li [2,7], Yimei Zhu [2], Mingzhao Liu [2], Peter V. Sushko [8, *], Andrew L. Walter [1, *], and Andi M. Barbour [1, *]

[1] National Synchrotron Light Source II, Brookhaven National Laboratory, Upton, New York 11973, USA.

[2] The Condensed Matter Physics and Materials Science Department, Brookhaven National Laboratory, Upton, New York 11973, USA.

[3] Center for Functional Nanomaterials, Brookhaven National Laboratory, Upton, New York 11973, USA.

[4] IBM T. J. Watson Research Center, 1101 Kitchawan Road, Yorktown Heights, New York 10598, USA.

[5] Department of Materials Science and Chemical Engineering, Stony Brook University, Stony Brook, New York 11794, USA.

[6] Material Measurement Laboratory, National Institute of Standard and Technology, Gaithersburg, Maryland 20899, USA.

[7] Department of Physics and Astronomy, Stony Brook University, Stony Brook, New York 11794, USA.

[8] Physical and Computational Sciences Directorate, Pacific Northwest National Laboratory, Richland, Washington 99354, USA.

*Corresponding authors: Aswin kumar Anbalagan (aanbalaga1@bnl.gov); Peter V. Sushko (peter.sushko@pnnl.gov); Andrew L. Walter (awalter@bnl.gov); and Andi M. Barbour (abarbour@bnl.gov).

**Abstract**

Despite constituting a smaller fraction of the qubit's electromagnetic mode, surfaces and interfaces can exert significant influence as sources of high-loss tangents, which brings forward the need to reveal properties of these extended defects and identify routes to their control. Here, we examine the structure and composition of the metal-substrate interfacial layer that exists in Ta/sapphire-based superconducting films. Synchrotron-based X-ray reflectivity measurements of Ta films, commonly used in these qubits, reveal an unexplored interface layer at the metal-substrate interface. Scanning transmission electron microscopy and core-level electron energy loss spectroscopy identified an approximately 0.65 nm ± 0.05 nm thick intermixing layer at the metal-substrate interface containing Al, O, and Ta atoms. Density functional theory (DFT) modeling reveals that the structure and properties of the Ta/sapphire heterojunctions are determined by the oxygen content on the sapphire surface prior to Ta deposition, as discussed for the limiting cases of Ta films on the O-rich versus Al-rich $Al_2O_3$ (0001) surface. By using a multimodal approach, integrating various material characterization techniques and DFT modeling, we have gained deeper insights into the interface layer between the metal and substrate. This intermixing at the metal-substrate interface influences their thermodynamic stability and electronic behavior, which may affect qubit performance.

1. Introduction

In recent years, there has been significant progress in the development of platforms aimed at revolutionizing quantum computing hardware[1]. These advancements have the potential to address issues in fields like materials science and cryptography that would traditionally take decades to solve using classical computers. A breakthrough in quantum computing hardware has been the discovery of superconducting quantum circuits (SQC's)[2]. SQCs offer a promising approach for quantum computing due to their low error rates and scalability. However, the practical implementation of superconducting qubits in a quantum processor is hindered by their limited coherence lifetime ($T_1$), which, in turn, is thought to be primarily limited by microwave dielectric losses[3]. Various research groups are actively working to improve the $T_1$ lifetime through two main approaches: understanding the microscopic mechanisms to better control performance and exploring new material choices[4].

Among the candidate materials for the microwave resonator, such as metal nitrides[5], aluminum (Al)[6-7], niobium (Nb)[8], and tantalum (Ta) [9-10] only Ta based microwave resonators have enabled longer $T_1$ lifetimes (0.5 ms)[10] than other materials. Even though the coherence lifetime of Ta based superconducting qubits is higher compared to other materials, Ta still faces challenges, such as surface oxide formation as soon as they are removed from the ultra-high vacuum environment. This surface oxide layer is usually disordered, leading to two-level system (TLS) loss in superconducting transmon qubits, which in turn results in a dielectric loss in the qubit, thereby impacting the $T_1$ [11-12].

Recently various research groups (including ours) focused on a more detailed analysis of the Ta interfaces between the air and metal layer. This analysis necessitated the use of advanced characterization tools including synchrotron-based variable energy X-ray photoemission spectroscopy (VEXPS) and transmission electron microscopy (TEM) measurements to understand the growth mechanisms of this $Ta_2O_5$ layer and provide insights on the structure and composition of this $Ta_2O_5$/Ta (oxide-metal) interface[8,13-14]. These measurements provided further evidence that the oxide-metal interface in Ta transmon qubits is chemically less complex than those observed in Nb transmon qubits[8]. Even though Ta-based transmon qubits give a higher $T_1$ value compared to others, it is essential to find ways to improve these $T_1$ values further. One avenue receiving much attention is suppressing the formation of the surface oxide layer. The formation of native $Ta_2O_5$ layer is inevitable at room temperature due to Ta's negative Gibbs free energy ($\Delta G°_{f, 298 K} = -1904$ kJ · mol$^{-1}$)[15]. Similar kind of native growth of native oxide layer is also true for qubits based on other materials such as Nb.

One effective strategy for suppressing the surface oxide formation is the surface encapsulation of the Ta or Nb films before exposing them to the ambient air environment, with transition metals[16-17]. Bal *et al.* demonstrated that Ta-capped niobium qubit devices exhibited $T_1$ relaxation times 2 to 5 times longer than native Nb films[16]. In another recent work, Zhou *et al.* demonstrated that a 3 nm Mg capping layer on top of Ta films can suppress surface oxidation by 60% compared to uncapped Ta films[17]. However, all these works discuss the oxide formation at the metal-air (M-A) interface and its impact on the $T_1$ of a superconducting transmon qubit. There are also other interfaces, such as the substrate-air (S-A) and the metal-substrate (M-S) interfaces, which also likely give rise to TLS[18,19,20]. There has currently, to the authors' knowledge, been no experimental or theoretical study reported on the M-S interface in the case of Ta/sapphire based superconducting qubits. Understanding the linkages between the synthesis/processing conditions, the resulting structure, fundamental properties, and the performance of these qubits is necessary to enable significant advancements in quantum computing technology.

In this work, we investigated the extent of structural and compositional inhomogeneity at the M-S interface of a superconducting Ta film (30 nm) on c-plane sapphire. The interfacial chemical states and roughness of the Ta film were examined using variable-energy X-ray photoelectron spectroscopy (VEXPS) and X-ray reflectivity (XRR). The phase purity of the Ta film and the epitaxial relation of the film were evaluated by X-ray diffraction (XRD) and transmission electron microscopy (TEM) techniques, respectively. Chemical profile analysis via high-angle annular dark-field scanning transmission electron microscopy (HAADF-STEM) and electron energy loss spectroscopy (EELS) was performed to obtain depth profile information across the metal-substrate interface. Density function theory (DFT) modeling was used to determine the origin of the M-S interface structural ordering using different sapphire termination models. Combining the insight obtained from XRR and STEM experiments and DFT modeling, we reveal the presence of an unexplored layer at the M-S interface, which is formed by the intermixing of Ta, Al, and O atoms from the metal and substrate layers. These findings offer valuable insights into controlling the structure and composition of the substrate-metal interface, thus paving the possible way to longer qubit coherence times.

## 2. Results and Discussions

A comprehensive description of the Ta thin film deposition process, the characterization methods employed in this study, and the DFT modeling approach can be found in the supporting information. XRR measurements were performed on Ta film using lab-based and synchrotron-based X-ray sources to determine the thickness, electron density, and roughness (**Figure S1**). The XRR of Ta films demonstrates a sharper interface in the Ta film, as indicated by the oscillation fringes up to a $Q_z$ value of 1.4 Å$^{-1}$ (where $Q_z = 4\pi\sin\theta/\lambda$, $\theta$, and $\lambda$ represent the Bragg angle and incident X-ray wavelength, respectively) in the case of synchrotron-based measurements compared to lab-based measurements. The poorer statistics at higher values of $Q_z$ in lab-based instruments (**Figure S1a**) are due to the lower X-ray beam flux and hence lower detector/energy resolution. In **Figure S1b**, the samples were measured at different φ (azimuthal) angles to determine any influence on the roughness of the Ta film due to possible surface terracing resulting from a surface miscut. Where φ represents the sample stage's rotation around the sample's normal axis. Upon comparison of the two phi angles, we find that φ = 0° has the slightly sharper interfaces.

The synchrotron-based XRR measurements are fitted with different models using the genX software[21] (**Figure 1**). **Figure 1a** depicts a simple fitting model (**Model 1**), consisting of a slab of sapphire with a layer of Ta and a native Ta oxide ($Ta_2O_5$) above it. However, the fitting is poor at higher $Q_z$ values. In **Figure 1b**, an interface layer of mixed suboxide was introduced between the M-A interface (*i.e.,* $Ta_2O_5$/sub-oxide interface/Ta metal) (**Model 2**). The parameters used for this suboxide interface model are based on the findings from our previous work[13]. There have been slight changes in the fitting parameters from this fitting model, but the fitting remained poor at higher $Q_z$ values. **Figure 1c** shows the XRR fitting results after incorporating an interface layer at the M-S region, which we refer to as the metal-suboxide-substrate interface (M-SO-S) (**Model 3**). This model is more realistic than models 1 and 2, as the fitting improved at higher Q values. We also considered another fitting model (**Figure 1d**), in which we combined Models 2 and 3, incorporating an interface suboxide layer at the M-A interface and an interface layer at the M-S interface (**Model 4**). This model provided similarly improved results compared to the model in **Figure 1c**. These findings provide evidence of a suboxide layer approximately 0.275 nm ± 0.003 nm thick exists at the M-SO-S interface, exhibiting an interfacial roughness of 0.035 nm ± 0.003 nm and 0.091 nm ± 0.001nm at the M-SO-S and SO-S, respectively. Here the presented error bars are generated by the genX software and represent a 5% increase to the fitting figure of merit. The fitting results for the data in **Figure 1c-d** are summarized in **Table 1**. Additionally, these measurements suggest the importance of obtaining a better S/N ratio at higher $Q_z$ values, especially $Q_z > 1.2$ Å$^{-1}$ in this case to observe this M-SO-S interface layer.

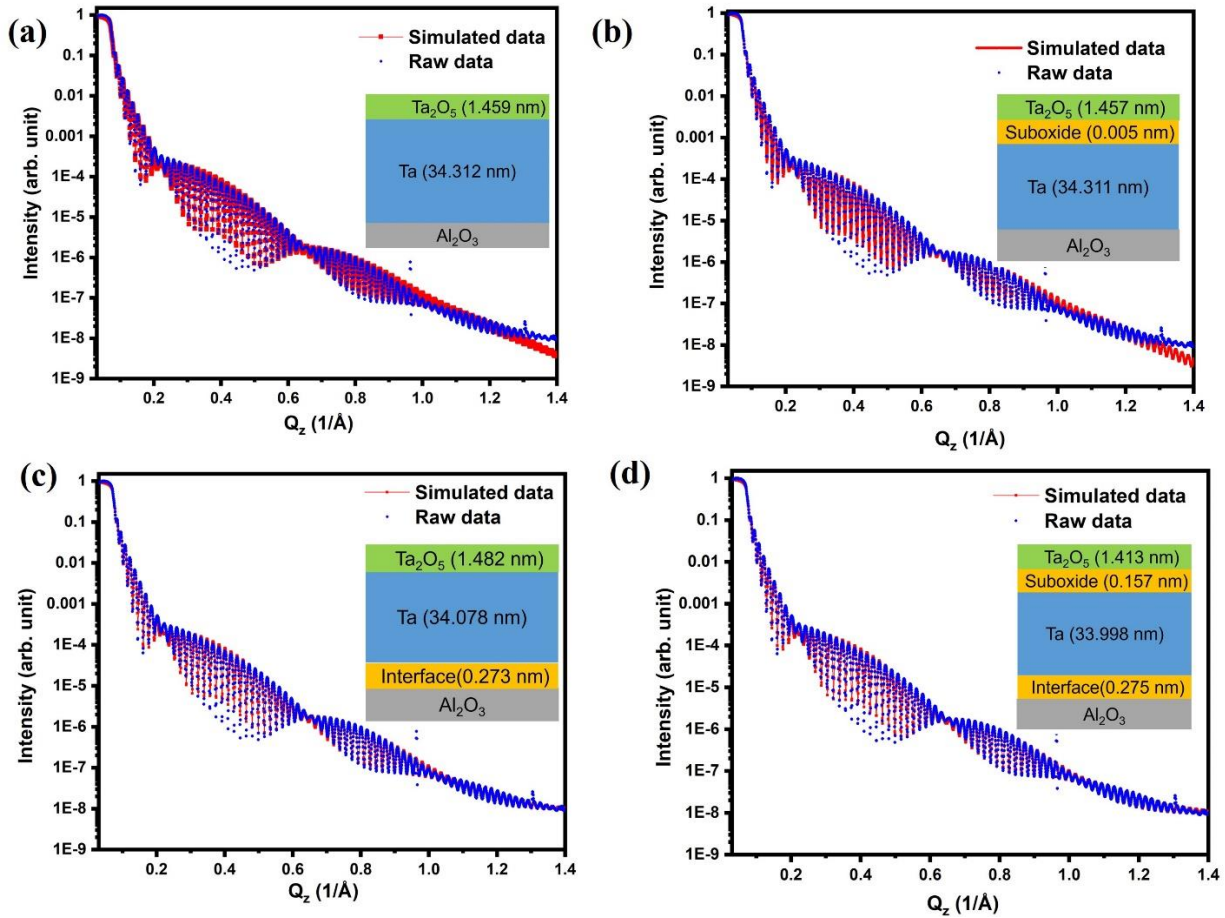

**Figure 1.** Fittings of the synchrotron X-ray reflectivity measurements for the BCC Ta film on sapphire substrate according to different layer models. (a) Model 1 neglecting transitional (buffer) layers, (b) Model 2 with an interfacial buffer layer between the Ta$_2$O$_5$ and metal layers, (c) Model 3 accounting for an interface layer between the substrate and metal layers, and (d) Model 4 with interface layers between both the $Ta_2O_5$/metal and substrate/metal layers. The inset shows the corresponding fitting values of thickness obtained.

Similar findings have been reported earlier in the case of Nb films (40 nm), where researchers demonstrated that their XRR measurements were best modeled by the addition of thin interface layers at the Nb-sapphire interface[22-23]. In a recent work, Satchell *et al.*[24] also confirmed that an additional interface layer is required to fully model the XRR measurements on a Nb (65 nm films)/sapphire substrate. The origin of this interface layer may correspond to either alloying or chemical reactions between them, leading to a possible formation of an NbAl layer. These previous reports on Nb films suggest that the thin, unexplored interface layer in Ta film could be due to the possible intermixing of Ta and sapphire atoms.

**Table 1.** Comparison of the Ta/sapphire heterojunction characteristics obtained by fitting XRR measurements using Models 3 and 4.

| Parameters | Model 3 (units in nm) | Model 4 (units in nm) |
|---|---|---|
| Thickness of $Ta_2O_5$ | 1.482 ± 0.020 | 1.413 ± 0.021 |
| Thickness of top suboxide layer | NA | 0.157 ± 0.017 |
| Thickness of Ta | 34.078 ± 0.012 | 33.998 ± 0.017 |
| Thickness of M-SO-S | 0.273 ± 0.004 | 0.275 ± 0.003 |
| Roughness of $Ta_2O_5$ | 0.249 ± 0.011 | 0.246 ± 0.011 |
| Roughness of top suboxide layer | NA | 0.226 ± 0.020 |
| Roughness of Ta | 0.216 ± 0.008 | 0.162 ± 0.009 |
| Roughness of M-SO-S | 0.043 ± 0.011 | 0.035 ± 0.003 |
| Roughness of SO-S | 0.092 ± 0.001 | 0.091 ± 0.001 |

XRD measurements were performed to determine the crystal structure of the sputtered Ta thin film on c-plane sapphire (0001). The out-of-plane XRD measurements (θ-2θ scan) shown in **Figure 2a,** confirmed that the Ta film exists in the pure BCC structure with a preferred orientation along the (222) direction, which is consistent with the orientation previously reported by Yao *et al*[25]. **Figure 2b** shows the cross-section TEM image of the Ta film with a thickness of about 34.0 nm ± 0.5 nm, consistent with the XRR fitting results. Additionally, a thin layer of approximately 4.0 nm ± 0.2 nm of amorphous native surface oxide exists on top of the Ta film. Here the measurement represents the average and standard deviation of ten unique measurements. The selected area electron diffraction (SAED) pattern **(Figure 2c)** illustrates that the TEM sample was measured along the [$\bar{1}\bar{1}0$] direction of Ta as the viewing axis and confirms the epitaxial relation between the Ta film and the sapphire substrate. The schematic in **Figure 2c** further illustrates the crystallographic orientation of the Ta film and the sapphire substrate. Laue pattern (**Figure S2)** further confirms the epitaxial relation between Ta film and sapphire.

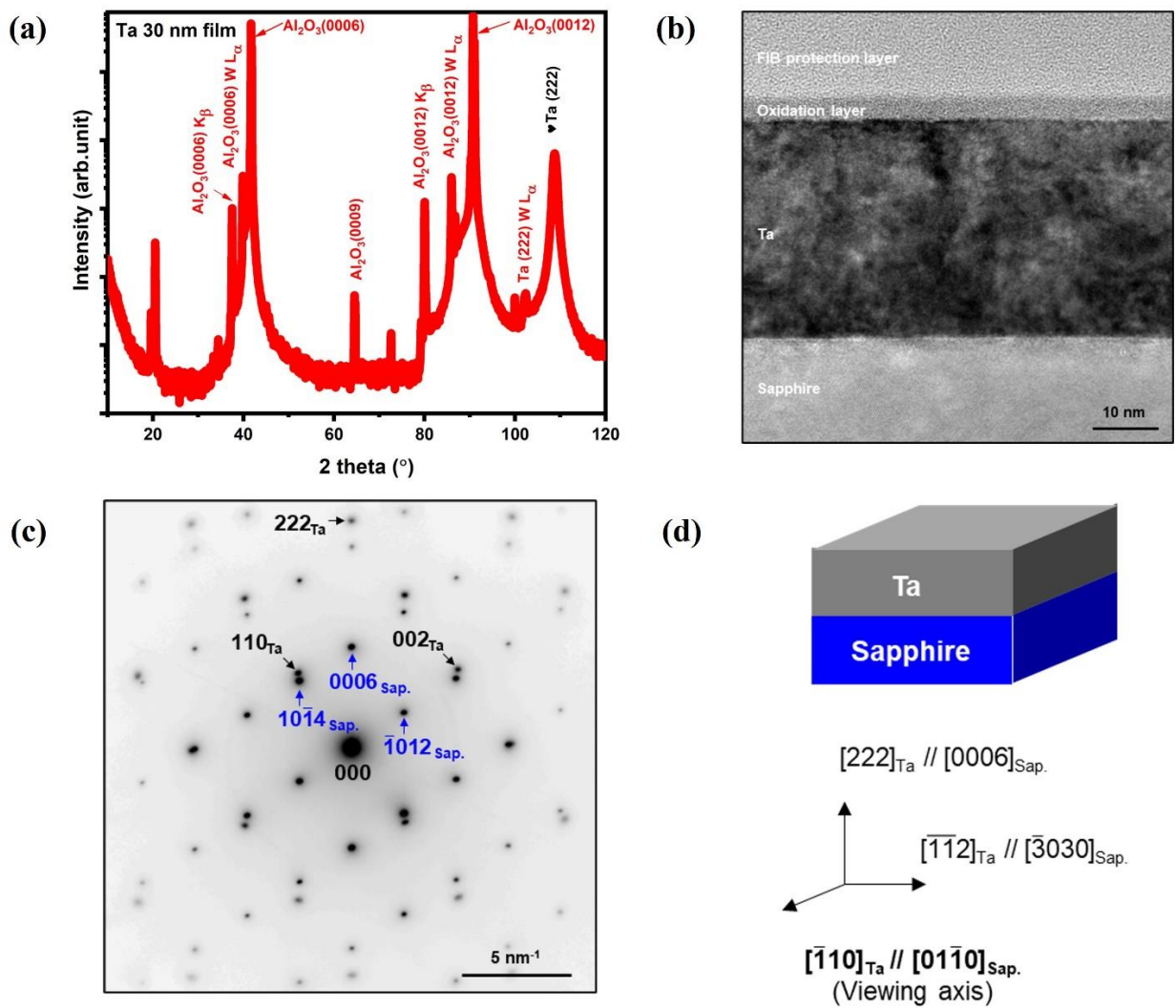

**Figure 2.** Analysis of the Ta film properties. (a) XRD pattern of an α-Ta film on C-plane sapphire, (b) cross-section TEM image of the Ta film, (c) the SAED pattern illustrating the epitaxial relationship between the Ta film and the substrate, and (d) the schematic representation of a 30 nm thick α-Ta film on c-plane sapphire substrate.

To correlate and confirm the existence of this interface layer at the M-S interface, chemical mapping at the atomic scale was employed using HAADF-STEM with EELS. **Figure 3** shows the HAADF-STEM false-colored mass contrast imaging analysis for atomic scale measurements in the M-S interface region. Although the Ta film exhibits high epitaxial quality with the sapphire substrate, the M-S interface appears broader rather than sharp. Additionally, we observe the mixing of Ta (green) and Al (red) in the interface layer. EELS scans were performed in multiple areas across the samples to understand this region better, as shown in **Figure S3a**. The energy of the Ta $L_3$ edge (9.8 keV) is not suitable for measurements due to its low S/N ratio[14]. Therefore, in this case, we performed EELS measurements in the energy region of 1400 eV to 1800 eV, which provides information about the Ta $M_{4,5}$ edge and Al K-edge. EELS analysis with a pixel size of about 0.05 nm on the broad interface layer confirmed a uniform intermixing of Ta, Al, and O atoms (**Figure S3b and S3c**). The thickness of this M-S interface layer was determined to be 0.65 nm ± 0.05 nm. This analysis is consistent with the results obtained from XRR measurements. These thickness values are close to the resolution limit for each method. The difference in the thickness values can be ascribed to the indistinct boundaries of the interface layer indicated by the HAADF-STEM measurements versus the slab-based modeling employed from the XRR fitting (0.275 nm ± 0.003 nm).

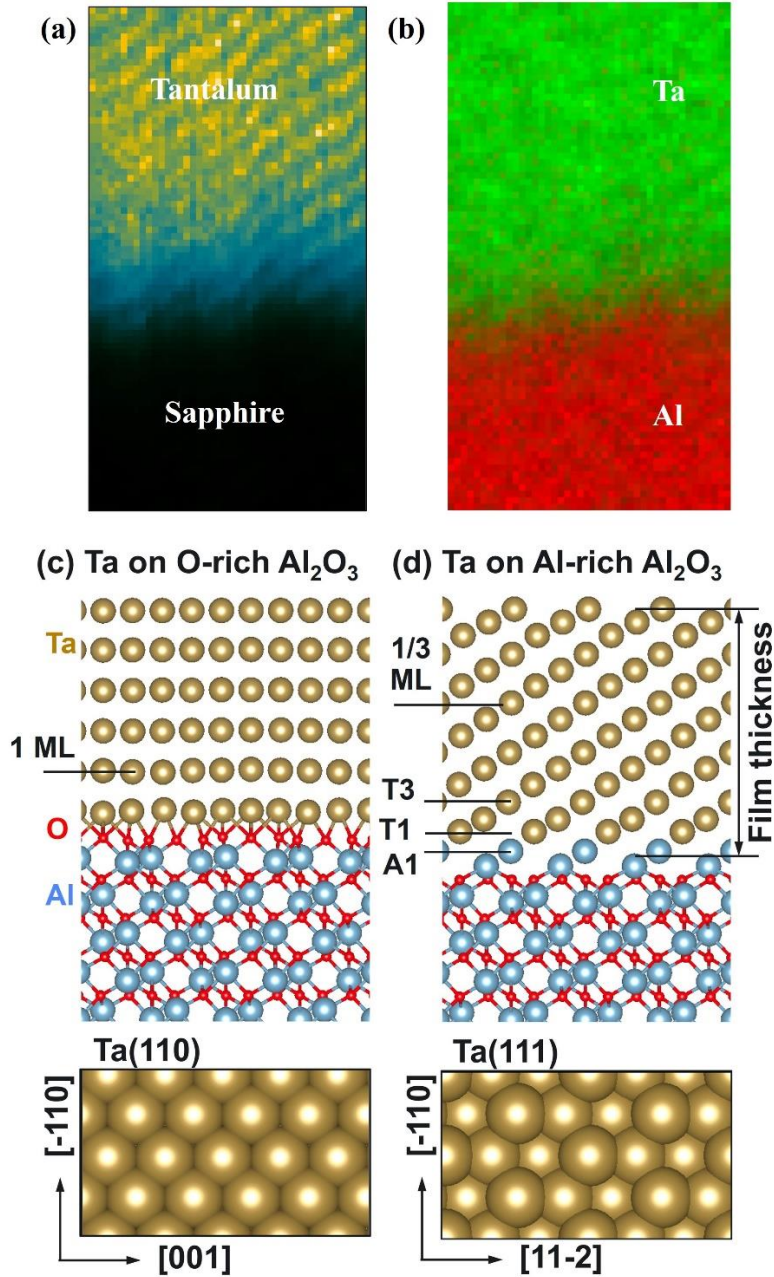

**Figure 3.** (a,b) HAADF-STEM false colored mass contrast imaging analysis of the Ta film on the metal/substrate interface and the intermixing of the Al-Ta shows the degree of interfacial broadening. (c,d) Structures of Ta/$Al_2O_3$ interfaces obtained using ab initio simulations: Ta(110) terminated film stabilized by Ta-O interactions (a) and Ta(111) terminated film formed as a continuation of the Al stacking sequence (b) on the O-rich and Al-rich substrates, respectively. The number density of Ta atoms per atomic plane in the O-rich case is approximately 3 times that in the Al-rich case.

In addition, the chemical profile of the Ta film across the surface region is quantified based on the VEXPS technique **(Figure S4)** to further correlate the surface oxide parameters (M-A interface) with the fitted XRR values. By varying the incident photon energy from low (690 eV) to high (4500 eV) (**see Figure S4 a-f**), the kinetic energy of emitted photoelectrons can be varied, thereby providing control over mean free paths and resulting in the measurement of photoelectrons at depths up to approximately 12 nm. The detailed description used for fitting the Ta 4f spectra, such as background subtraction, error calculations, and mixed suboxide features, can be found in our previous work[13]. **Figure S4 g-h** shows that beneath a thickness of 2 nm to 2.5 nm (from the top), most of the Ta film exists in the pure metallic state. This depth profiling result from VEXPS measurement is consistent with the previously published works on native Ta film[13,17]. The thickness obtained for $Ta_2O_5$ ($Ta^{5+}$) using VEXPS measurements is around 1.676 nm ± 0.034 nm, which is quite consistent with the XRR fitting. Moreover, this VEXPS data was collected sometime after the XRR experiments, which utilized a freshly deposited Ta film. Given this and our previous VEXPS results of freshly deposited films, we can also confirm the self-passivating nature of the oxide layer formed at the M-A interface.

To further our insight into the connection between the structure, stability, and electronic properties of $Ta/Al_2O_3$ heterojunctions, DFT modeling was utilized. The two types of $Ta/Al_2O_3$ interfaces formed by depositing Ta metal on the O-rich and Al-rich $Al_2O_3$ (0001) surfaces, respectively, are shown in **Figure 3c and 3d**. In the case of O-rich $Al_2O_3$ (0001), the terminating surfaces contain the full oxygen bilayer, which corresponds to the average formal charge state of the surface oxygen species of $O^-$. Upon relaxation, this surface undergoes disproportionation of the electron charge between the surface oxygen species resulting in the formation of peroxy ($O_2^{2-}$) and superoxide ($O_2^-$) ions in addition to $O^{2-}$. A single monolayer of Ta metal deposited on such a surface provides enough electrons to fully complete the O 2p shells of all surface oxygen species. The resulting Ta ions are distributed over the surface to form a (110) surface of the BCC Ta lattice, as shown in **Figure 3c.** The Ta film is subjected to 1.7% tensile strain in the [-110] direction. However, there is no obvious lattice match along the [001] direction. Inspection of the STEM images of the Ta/$Al_2O_3$ interfaces[9] suggests that the periodic structure of the interface is determined by the 7:8 ratio of the Ta to $Al_2O_3$ lattice spacings in the <-112> and <11-20> directions, respectively. Due to the constraints of the periodic model approach, we considered the supercells with the ratios of 5:6 and 8:9 (not shown), which correspond to approximately 4.2 % tensile and 1.4 % compressive strains of Ta film relative to the experimentally observed systems and result in similar structures.

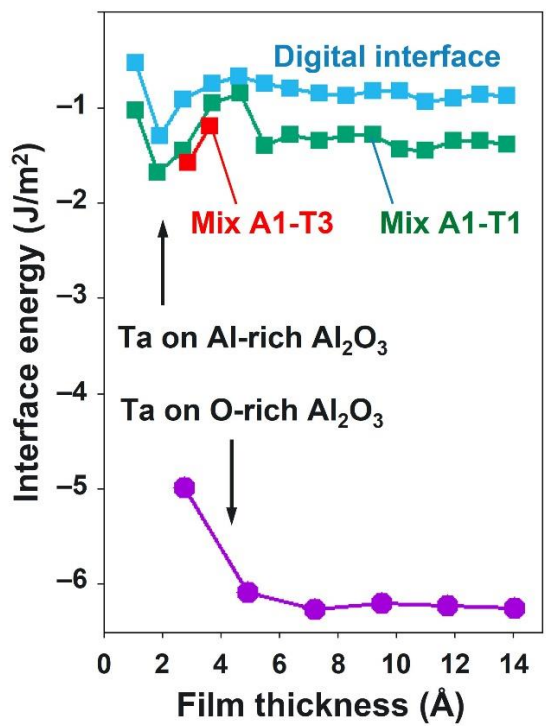

**Figure 4.** Interface energy as a function of Ta film thickness for the digital and intermixed configurations formed by swapping Al and Ta atoms in planes A1, T1, and T3 indicated in (Figure 3d).

In the case of the Al-rich surface, atoms of the first Ta monolayer (T1) were found to be most stable at the hollow sites of the Al bilayer. Accordingly, the first Ta plane has the same number density and atomic arrangement as the topmost Al plane, resulting in a seamless Ta/Al interface. Additional Ta atoms also occupy the hollow sites, thus forming a BCC Ta film terminated with the Ta (111) surface. This structure (**Figure 3d**) exhibits slight variations in the calculated interplane distances for the topmost Al–Al (0.783 Å), interfacial Al–Ta (1.163 Å), and Ta–Ta (0.898 Å on average) planes. Based on the experimental lattice parameters of the bulk $Al_2O_3$ and Ta, the film is subjected to an approximately 1.7% uniform tensile strain. A similar strain value (approximately 2.1 %) is obtained if the computed lattice parameters are used instead.

To evaluate the thermodynamic stabilities of these Ta films on the $Al_2O_3$ substrate (**Figure 4**), we first compared the surface energies of the relaxed (110) and (111) surfaces (2.33 $J/m^2$ and 2.74 $J/m^2$, respectively) and the energies required to strain Ta slabs terminated with these surfaces to match the supercell parameters of the $Al_2O_3$ substrate: approximately 25 meV per atom in both cases. Then, the interfacial energies of the films were calculated relative to the energies of pure O-rich and Al-rich alumina surfaces and the cohesive energy of the bulk Ta and corrected for the contributions due to the substrate-induced strain and energy of the exposed surface, resulting in –6.25 $J/m^2$ and –0.88 $J/m^2$, respectively. These values correspond to the limiting cases of the fully oxidized and fully reduced surfaces. In general, interfacial energies and structures of these interfaces are defined by the oxygen content and spatial distribution of these oxygen species over the surface.

The charge distribution for both types of the $Ta/Al_2O_3$ structures is shown in **Figure 5 & S5**. In the case of the O-rich $Al_2O_3$ substrate, Ta atoms of the first monolayer donate their electrons to the outermost oxygen plane, thus creating an interfacial dipole in the out-of-plane direction (**Figure 5a**). This charge redistribution is short ranged as the Ta atoms beyond the first interfacial plane are neutral. For comparison, in the case of the Al-rich $Al_2O_3$ substrate, the electron charge is transferred from the outermost Al plane to the second from the interface Ta plane, resulting in the interfacial dipole opposite to that found for the O-rich case (**Figure 5b**).

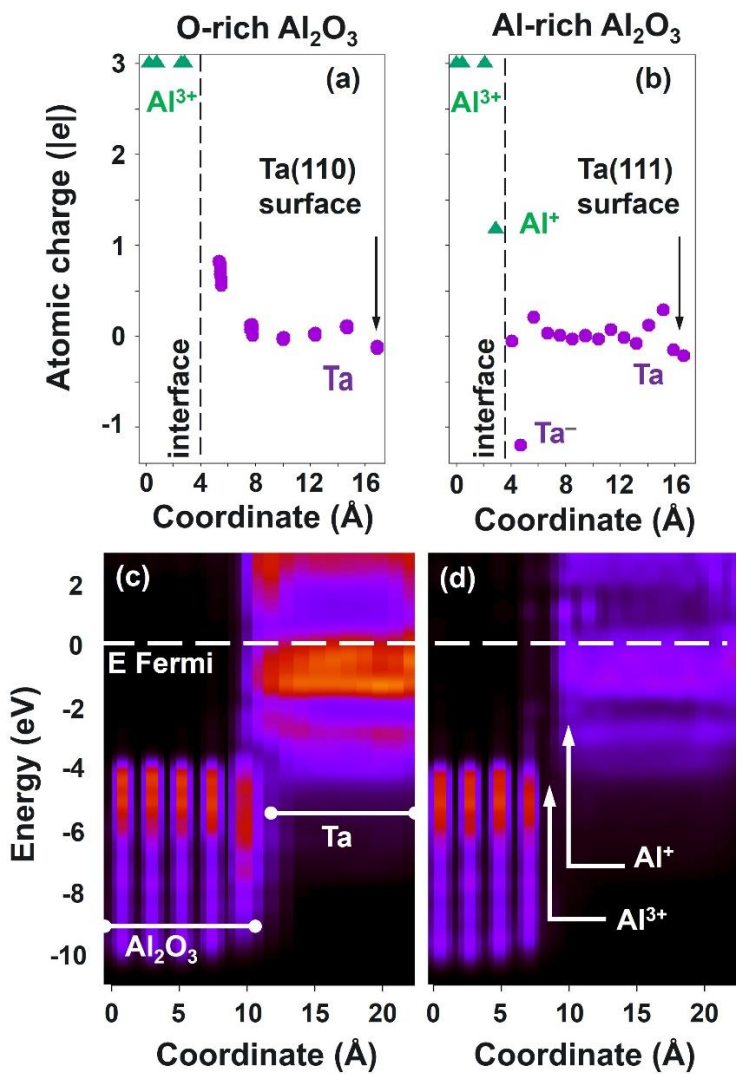

**Figure 5.** (a,b) Charge distributions in selected Ta/Al$_2$O$_3$ systems represented by layer-averaged Bader atomic charges for Ta on the O-rich (a) and Al-rich (b) Al$_2$O$_3$. (c,d) Heatmap representations of one-electron DOS calculated for the Ta on the O-rich (c) and Al-rich (d) Al$_2$O$_3$ systems and projected on the Al, O, and Ta atomic planes. The energy scales are aligned so that Fermi energy is at 0 eV. The difference in the Ta DOS intensity in (c) and (d) is due to different Ta stacking sequences along the out-of-plane direction.

To assess the width of the region affected by such electron redistribution, we calculated the interfacial energy as a function of the film thickness (**Figure 4**). In the O-rich case, this energy is the largest for the first Ta monolayer and nearly constant after that, consistently with the charge distribution. In contrast, in the Al-rich case, the interfacial energy shows noticeably larger variations with the film thickness indicating its sensitivity to the details of the atomic positions across the film. Given that Al metal cohesive energy (3.39 eV/atom) is much smaller than that of the bulk Ta (8.10 eV/atom)[26], it can be expected that the outer plane of Al (A1 in **Figure 3d**) dissolves into the Ta films at the initial stages of the film deposition. In particular, we found that deposition of the first 1/3 ML of Ta atoms on the Al-rich surface results in a spontaneous in-diffusion of these atoms so that they occupy the space between two Al planes (**see Figure S6**).

To assess the effect of such Ta-Al intermixing, we considered several configurations in which Al atoms of the outermost substrate plane (denoted as A1) were swapped with Ta atoms in the 1$^{st}$, 2$^{nd}$, or 3$^{rd}$ place from the interface (denoted as T1, T2, and T3, respectively). The A1-T1 swap was found to stabilize the Ta film by as much as 0.5 J/m$^2$ for all film thicknesses and the A1-T3 swap has an even larger stabilizing effect for up to 4/3 ML Ta coverage (**Figure 4**), while the A1-T2 swap had a destabilizing effect only (not shown). These calculations suggest that excess Al indeed dissolves into the Ta film forming a buffer layer, consistent with the experimentally observed transition layer. Furthermore, charge distribution for the intermixed configurations **(see Figure S5)** in the case of the thick film shows the sequence of positively and negatively charged layers formed by the Al and Ta species respectively which stretches at least 0.6 nm into the Ta film, which qualitatively agrees with the width of the buffer layer estimated from the experimental observations.

Finally, to analyze the electronic structure of the Ta/Al$_2$O$_3$ interfaces, we plotted the one-electron density of states projected on the atomic planes for selected cases (**Figure 5 c, d & S7**). In the case of the O-rich substrate, the interface is electronically abrupt and expected from the well-ordered structure of the interface (**Figure 5c**). In the case of the ideal Ta on the Al-rich Al$_2$O$_3$, there is a noticeable presence of metal states in the Al$_2$O$_3$ band gap region near this interface (**Figure 5d**). As this interface is stabilized via Al-Ta intermixing, the DOS near Al in the Ta film is severely depleted even though the Ta-Al film is nominally metallic. We attribute this effect to the charge disproportionation between the Al and Ta species resulting in positively charged Al species as indicated in **Figure 5**. Overall, the DFT modeling, corroborated by HAADF-STEM analysis, confirm that our Ta films are grown on Al-rich Al$_2$O$_3$ surface.

With well-characterized Ta interfaces and theoretically predicted interface structure, which is usually metallic but has charge depleted Al, it remains to be understood how this M-S interface layer can affect the physics governing qubit performance. To address this, low-temperature transport measurements were conducted to determine the effects on the superconducting properties of the Ta film due to the existence of an interface layer between the metal and substrate layers. **Figure 6a** shows the resistivity measured as a function of temperature for the Ta film, which has a superconducting transition temperature $T_c$ of 3.84 K and mid-point transition $T_c$ of 3.93 K, with a residual resistivity ratio (RRR) value ($\rho$ (4 K)/ $\rho$ (300 K)) of 4.98. This $T_c$ value is slightly lower than those reported in previous works for a pure BCC structure Ta film ($T_c$ = 4.0 K to 4.4 K)[17,25,27]. However, several previous studies have also reported T$_c$ values of BCC structure Ta films to be less than 4 K[28-30]. **Table S1** summarizes some of the previous studies performed on Ta-based superconducting films on various sapphire and silicon substrates. **Figures 6b and 6c** show the AC magnetic susceptibility measurements of the Ta film under a drive amplitude of 1 Oe. The magnetic field was applied parallel to the film's surface. The mid-point $T_c$ of the $\chi'$ and the peak location of $\chi''$ is at 3.86 K, similar to the $T_c$ obtained from the resistivity measurement. The lower $T_c$ could be attributed to the presence of the M-S interface layer that we experimentally observed and confirmed its possible existence through DFT modeling. Alternatively, it could be due to the defects present in the films. This, in turn, resulted in a lower RRR value, indicating the film exhibits type-II superconductivity.

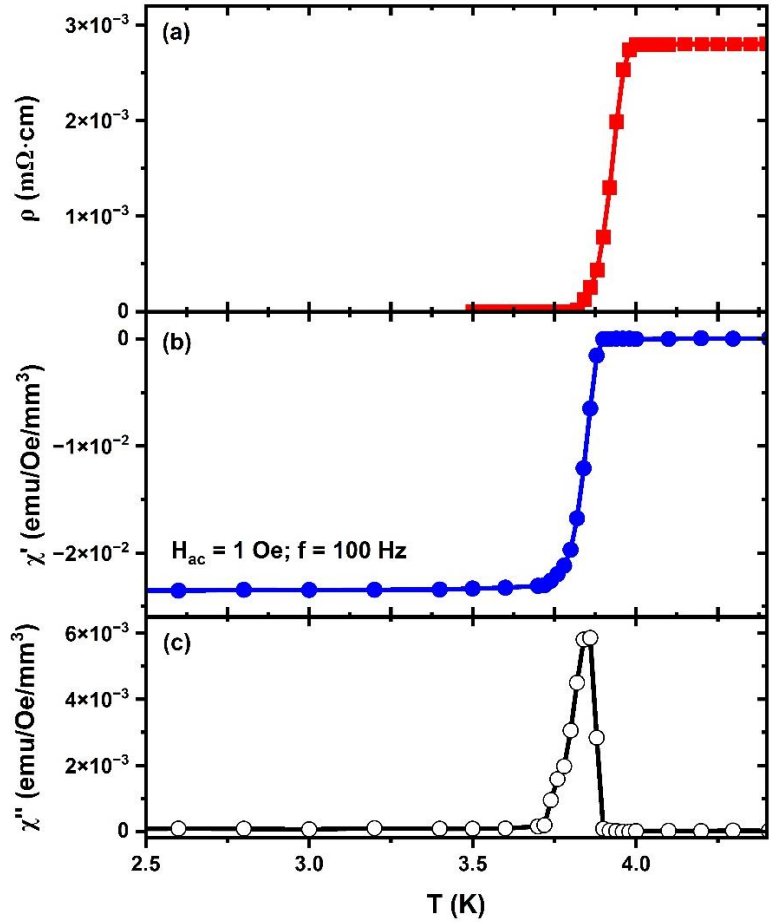

**Figure 6.** (a) Temperature dependent resistivity of the 30 nm Ta film. (b) and (c) The real and imaginary component of ac magnetic susceptibility of the same film under a drive amplitude of 1 Oe and a frequency of 100 Hz.

This multimodal study provides concrete evidence of an interface layer between M-S layer, identified using XRR and HAADF-STEM techniques. Additionally, DFT modeling reveals that the atomic and electronic properties of Ta/$Al_2O_3$ heterojunctions are significantly influenced by the surface composition of $Al_2O_3$. Specifically, O-rich surfaces promote the formation of peroxide or superoxide ions, while Al-rich surfaces result in considerable Ta-Al intermixing at the interface. These variations critically affect the thermodynamic stability and electronic characteristics of the heterojunctions, potentially affecting the $T_1$ time of qubits.

Generally, material loss research is currently focused on three factors in order to improve qubit lifetimes: dielectric properties of the substrates, Josephson junction losses, and interface losses[31]. Previous 3D cavity measurements have shown that the loss tangent of a 5 nm thick oxide is approximately 0.1, which exceeds the losses at the metal/substrate interface. However, Wisbey *et al.* found that TLS loss is influenced by both the M-S interface and surface roughness[32]. Looking forward, it is not unreasonable to suggest that studies involving bulk loss in resonators material will become more important when M-S and M-A interfaces cannot be further engineered to have significant impact on TLS loss. In that light, the presented DFT findings elucidate the growth mechanism for Ta (110) or Ta (111) orientations on different $Al_2O_3$ (0001) terminated surfaces. However, the effects of mixed surface terminations of sapphire on the crystallography of the deposited Ta film remain are not yet fully understood [33]. Specifically, the presence of mixed crystalline habitats and the consistency of the imposed strain across bulk Ta films in the 150 nm to 300 nm thickness range need further clarification. Additionally, more investigation is required to understand the relaxation process of this strain to its nominal value and its impact on the film's morphology and roughness[9]. The current study indicates that although the M-S interface may not directly contribute to TLS, the M-S interface plays a crucial role in determining the characteristics of the resulting film. We have demonstrated several experimental methods to investigate this region, and our DFT results suggest that previously reported record-breaking lifetimes are associated with O-terminated sapphire.

3. Conclusion

This study demonstrates that the epitaxial growth of a Ta (111) thin film on a c-plane (0001) sapphire leads to the formation of an interface layer at the metal-substrate interface and that the structure of this layer is potentially controlled at the atomic level by tuning the sapphire surface termination. Using a combination of synchrotron X-ray reflectivity and scanning transmission electron microscopy, we investigated the quality of the metal-substrate interface layer. XRR revealed the unexplored interface layer at the M-S interface is about 0.275 nm ± 0.003 nm. HAADF-STEM measurements confirmed this interfacial layer, indicating a thickness of 0.65 nm ± 0.05 nm. The difference in thickness can be attributed to the indistinct boundaries of the interface layer observed in HAADF-STEM. The structure, stability, and electronic properties of $Ta/Al_2O_3$ heterojunctions examined by DFT modeling further confirmed the possible intermixing at the M-S interface. This multimodal approach, employing various material characterization techniques, enhances our understanding of the interface layer between the metal and substrate. These insights could pave the way for controlling this interface layer and hence improving the coherence time of superconducting qubits.

**Authors contribution**

A.k.A., A.L.W., and A.B., conceived the idea and designed the experiments. A.k.A., A.B., J.J.S., and V.S. performed synchrotron-based XRR experiments. M.L., and C.Z. were responsible for material synthesis and J.Y., and Q. Li for low-temperature transport measurements. R.C., J.M., and K.K. conducted electron microscopy experiments. A.k.A. and D.N. performed lab-based XRR measurements. P.V.S. carried out theoretical calculations. A.k.A., C.Z., and C.W. performed synchrotron-based XPS experiments. S.L.H., N.P. L., Y.Z., M.L., P.V.S., A.L.W., and A.B. supervised the project. A.k.A., P.V.S., A.L.W., and A.B. drafted the manuscript. All authors participated in editing the manuscript.

**Conflict of Interest**

The authors declare no conflicts of interest.

**Acknowledgments**

This project received support from the U.S. Department of Energy (DOE), Co-design Center for Quantum Advantage ($C^2QA$), under contract number DE-SC0012704. The research used synchrotron X-ray resources of the National Light Source II; a U.S. DOE Office of Science User Facility operated for the DOE Office of Science by Brookhaven National Laboratory (BNL) under DOE contract DE-SC0012704. In particular, it used National Institute of Standards and Technology (NIST) beamline 07-ID-2 (SST-2) and the & International Business Machines Corporation (IBM) operated endstation at the NIST 06-BMM) beamline. Materials Synthesis & Characterization, and Electron Microscopy facilities of the Center for Functional Nanomaterials (CFN), DOE Office of Science Facilities at BNL, where also used under contract number DE-SC0012704. The authors also would like to acknowledge the use of facilities within the Electron Microscopy and Nanostructure Group, and Advanced Energy Materials Group (J.Y., Q. Li), the Department of Condensed Matter Physics & Materials Science at BNL, supported by DOE, Office of Basic Energy Sciences, Division of Materials Sciences and Engineering, under Contract No. DE-SC0012704. Computational modeling at the Pacific Northwest National Laboratory (PNNL) was supported by C$^2$QA (BES, PNNL FWP 76274). This research used resources of the National Energy Research Scientific Computing Center; a DOE Office of Science User Facility supported by the Office of Science of the U.S. DOE under Contract No. DE-AC02-05CH11231 using NERSC award BES-ERCAP0028497.We would also like to acknowledge Nathalie de Leon for the fruitful discussions.**Disclaimer**

Certain commercial equipment, instruments, or materials are identified in this paper in order to specify the experimental procedure adequately, and do not represent an endorsement by the National Institute of Standards and Technology.**References**

(1) Arute, F.; Arya, K.; Babbush, R.; Bacon, D.; Bardin, J. C.; Barends, R.; Biswas, R.; Boixo, S.; Brandao, F. G. S. L.; Buell, D. A.; et al. Quantum supremacy using a programmable superconducting processor. *Nature* **2019**, *574* (7779), 505-510. DOI: 10.1038/s41586-019-1666-5.
(2) Gyenis, A.; Di Paolo, A.; Koch, J.; Blais, A.; Houck, A. A.; Schuster, D. I. Moving beyond the Transmon: Noise-Protected Superconducting Quantum Circuits. *PRX Quantum* **2021**, *2* (3), 030101. DOI: 10.1103/PRXQuantum.2.030101.24

superconducting transmon qubits with coherence times exceeding 0.3 milliseconds. *Nat. Commun.* **2021**, *12* (1), 1779. DOI: 10.1038/s41467-021-22030-5.

(10) Wang, C.; Li, X.; Xu, H.; Li, Z.; Wang, J.; Yang, Z.; Mi, Z.; Liang, X.; Su, T.; Yang, C.; et al. Towards practical quantum computers: transmon qubit with a lifetime approaching 0.5 milliseconds. *npj Quantum Inf.* **2022**, *8* (1), 3. DOI: 10.1038/s41534-021-00510-2.

(11) Siddiqi, I. Engineering high-coherence superconducting qubits. *Nat. Rev. Mater.* **2021**, *6* (10), 875-891. DOI: 10.1038/s41578-021-00370-4.

(12) Crowley, K. D.; McLellan, R. A.; Dutta, A.; Shumiya, N.; Place, A. P. M.; Le, X. H.; Gang, Y.; Madhavan, T.; Bland, M. P.; Chang, R.; et al. Disentangling Losses in Tantalum Superconducting Circuits. *Phys. Rev. X* **2023**, *13* (4), 041005. DOI: 10.1103/PhysRevX.13.041005.

(13) McLellan, R. A.; Dutta, A.; Zhou, C.; Jia, Y.; Weiland, C.; Gui, X.; Place, A. P. M.; Crowley, K. D.; Le, X. H.; Madhavan, T.; et al. Chemical Profiles of the Oxides on Tantalum in State of the Art Superconducting Circuits. *Adv. Sci.* **2023**, *10* (21), 2300921. DOI: https://doi.org/10.1002/advs.202300921.

(14) Mun, J.; Sushko, P. V.; Brass, E.; Zhou, C.; Kisslinger, K.; Qu, X.; Liu, M.; Zhu, Y. Probing Oxidation-Driven Amorphized Surfaces in a Ta(110) Film for Superconducting Qubit. *ACS Nano* **2024**, *18* (1), 1126-1136. DOI: 10.1021/acsnano.3c10740.

(15) Gebauer, J.; Franke, P.; Seifert, H. J. Thermodynamic Evaluation of the system Ta–O and Preliminary Assessment of the Systems Al–Nb–O and Al–Ta–O. *Adv. Eng. Mater.* **2022**, *24* (8), 2200162. DOI: https://doi.org/10.1002/adem.202200162.26

(32) Wisbey, D. S.; Gao, J.; Vissers, M. R.; da Silva, F. C. S. ; Kline, J. S.; Vale, L.; Pappas, D. P. J. Appl. Phys. (2010) 108, 093918.

(33) Gnanarajan, G.; Lam, S.K.H.; Bendavid, A. Coexistence of epitaxial Ta (111) and Ta(110) oriented magnetron sputtered thin film on c-cut sapphire. *J. Vac. Sci. Technol. A* (2010) 28, 175–181.

**Supporting information for**

**Revealing the Origin and Nature of the Buried Metal-Substrate Interface Layer in Ta/Sapphire Superconducting Films**

Aswin kumar Anbalagan [1, *], Rebecca Cummings [2], Chenyu Zhou [3], Junsik Mun [2,3], Vesna Stanic [4], Jean Jordan-Sweet [4], Juntao Yao [2,5], Kim Kisslinger [3], Conan Weiland [6], Dmytro Nykypanchuk [3], Steven L. Hulbert [1], Qiang Li [2,7], Yimei Zhu [2], Mingzhao Liu [2], Peter V. Sushko [8, *], Andrew L. Walter [1, *], and Andi M. Barbour [1, *]

[1] National Synchrotron Light Source II, Brookhaven National Laboratory, Upton, New York 11973, USA.

[2] The Condensed Matter Physics and Materials Science Department, Brookhaven National Laboratory, Upton, New York 11973, USA.

[3] Center for Functional Nanomaterials, Brookhaven National Laboratory, Upton, New York 11973, USA.

[4] IBM T. J. Watson Research Center, 1101 Kitchawan Road, Yorktown Heights, New York 10598, USA.

[5] Department of Materials Science and Chemical Engineering, Stony Brook University, Stony Brook, New York 11794, USA.

[6] Material Measurement Laboratory, National Institute of Standard and Technology, Gaithersburg, Maryland 20899, USA.

[7] Department of Physics and Astronomy, Stony Brook University, Stony Brook, New York 11794, USA.

[8] Physical and Computational Sciences Directorate, Pacific Northwest National Laboratory, Richland, Washington 99354, USA.

*Corresponding authors: Aswin kumar Anbalagan (aanbalaga1@bnl.gov); Peter V. Sushko (peter.sushko@pnnl.gov); Andrew L. Walter (awalter@bnl.gov); and Andi M. Barbour (abarbour@bnl.gov).

1. Experimental

1.1 Thin film preparation

Tantalum thin film sample was prepared by RF magnetron sputtering on a 2-inch C-cut sapphire (0001) (CrysTec, single-side polished) at a substrate temperature of 750 °C in an AJA Orion Sputtering System equipped with A315-UHV sputtering sources. The base pressure of the sputtering chamber was maintained at $10^{-6}$ Pa. Before loading the sapphire wafer into the deposition chamber, it was cleaned with a piranha solution made by mixing a commercial 30% hydrogen peroxide solution and 98% sulfuric acid in a volumetric ratio of 1:2, followed by rinsing with DI water. The Ta target (99.98 % purity) was procured from AJA International, Inc. The sputtering was performed at a working pressure of 1.6 Pa, with Research Grade Ar (Airgas) as plasma gas and a sputtering power of 100 W (RF).

1.2 Materials characterization

Synchrotron-based X-ray reflectivity (XRR) measurements were performed using 8.6 keV energy on the 06-BMM beamline operated by the National Institute of Standards and Technology (NIST) at the National Synchrotron Light Source (NSLS-II), Brookhaven National Laboratory (BNL). For these measurements, a Si (111) monochromator was used, which has an energy resolution of 1.3 × $10^{-4}$ ΔE/E. The exit slit size at the Mythen detector was 0.12 mm (V) × 1 mm (H) before the reflected signal was collected. Background subtraction for the reflectivity measurements were performed by collecting pixel zero (60 pixels), integrating it from both sides, applying Gaussian smoothing (10 points), and then averaging 2 curves and subtracting them. Lab-based X-ray diffraction (XRD) and XRR measurements were collected on a Rigaku SmartLab II X-ray diffractometer using Cu Kα radiation. The fitting of these reflectivity measurements was performed using the genX software[1]. Transmission electron microscopy (TEM), high-angle annular dark-field (HAADF) with scanning tunning electron microscopy (HAADF-STEM), and electron energy loss spectroscopy (EELS) measurements were carried out on a JEOLARM-200F, with the sample prepared by focused ion beam (FIB) lift-off technique utilizing a dual beam SEM/FIB microscope (FEI Helios). Variable energy X-ray photoemission spectroscopy (VEXPS) measurements were performed at the SST-2 beamline operated by the NIST at NSLS-II, BNL. The energy of the incident X-ray beam was varied from 690 eV (soft X-ray regime) to 4500 eV (tender X-ray regime). During the measurements, the beam's incident angle to the sample was kept at 10° and the takeoff angle of the photoelectrons to the detector was 80°. A gold standard was used for energy calibration. Transport measurements were performed by the four-probe inline method in a 14 Tesla Quantum Design physical property measurement system (PPMS). The magnetic ac susceptibility measurements were measured in a Quantum Design magnetic properties measurement system (MPMS) with a SQUID (superconducting quantum interference device) magnetometer.

### 1.3 Density functional theory (DFT) modeling

The structure, stability, and electronic properties of Ta/$Al_2O_3$ heterojunctions were analyzed using the periodic slab model and two types of terminating $Al_2O_3$ (0001) surfaces: Al-rich and O-rich ones. Ta films, commensurate with either Al-rich or O-rich terminations of $Al_2O_3$ (0001), were deposited on both sides of the $Al_2O_3$ slab to form equivalent terminating Ta surfaces. The calculations were performed using the VASP package [2,3] and Perdew-Burke-Ernzerhof (PBE) [4] exchange-correlation functional. The projector-augmented-wave potentials were used to approximate the effect of the core electrons [5].

The plane-wave basis set cutoff was set to 500 eV. The total energy convergence criterion was set to $10^{-5}$ eV. Bader charge population analysis was used to analyze the charge density redistribution [6,7]. The lateral supercell parameters were based on the bulk $Al_2O_3$ hexagonal lattice parameter (4.80813 Å) pre-calculated using the crystallographic cell and 5×5×2 k-mesh. The total energies were minimized with respect to all internal coordinates. The adsorption energies of Ta atoms were calculated relative to the Ta bulk cohesive energy. One-electron densities of states (DOS) were smeared out by convoluting band energies with Gaussian functions with the full width at a half-maximum of 0.1 eV.

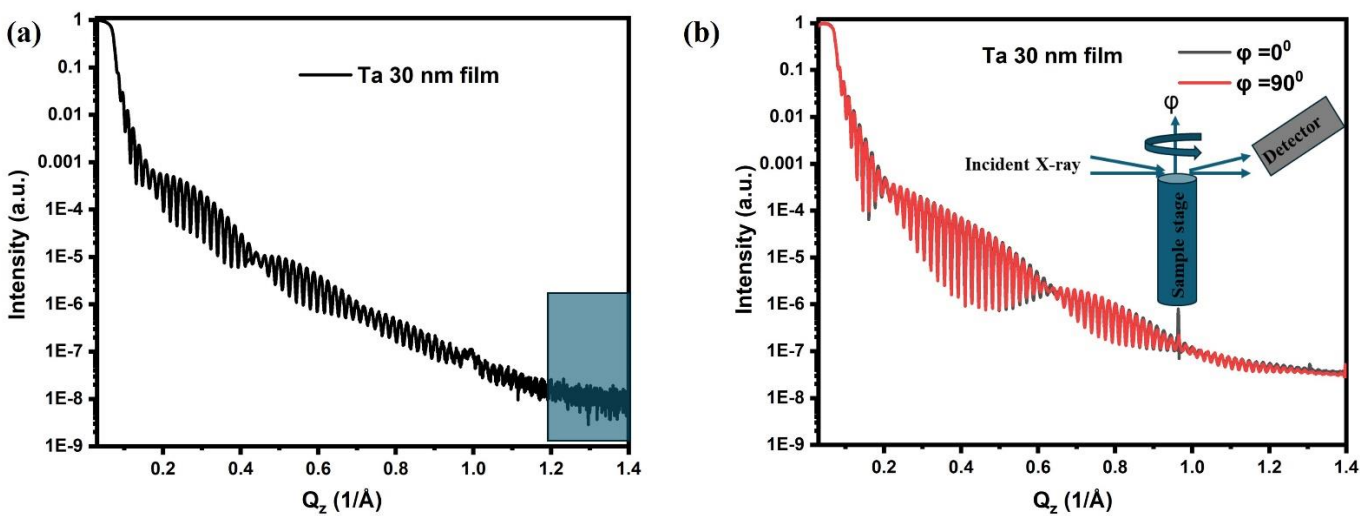

**Figure S1.** X-ray reflectivity of Ta film using (a) lab-based and (b) synchrotron-based tools. The shaded area in the lab-based XRR plot highlights the crucial range of $Q_z$ values responsible for determining the information about the metal-substrate interface of the Ta film.

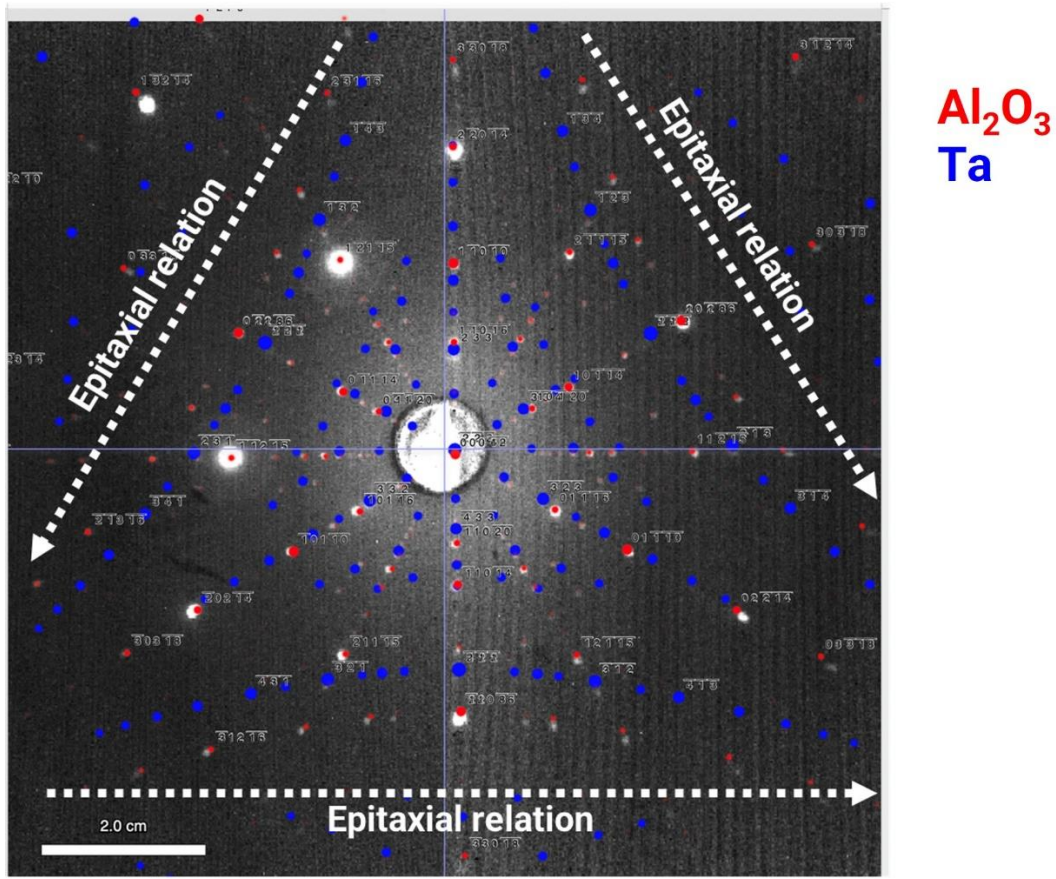

**Figure S2.** The Laue pattern illustrates the epitaxial relationship between the Ta film and the substrate. For phi=0 in the vertical scattering geometry, the incident X-rays come from the bottom (or top) of the page.

EELS scans were performed in multiple areas across the samples, as shown in **Figure S3a**, to better understand this region. **Figure S3b, c** shows the EELS scan of spots 1 and spot 2 measured in the energy region of 1400-1800 eV. These EELS measurements reveal the co-existence of both Ta and Al atoms at all measured spots, indicating the broad interface layer is uniform across the metal-substrate layer.

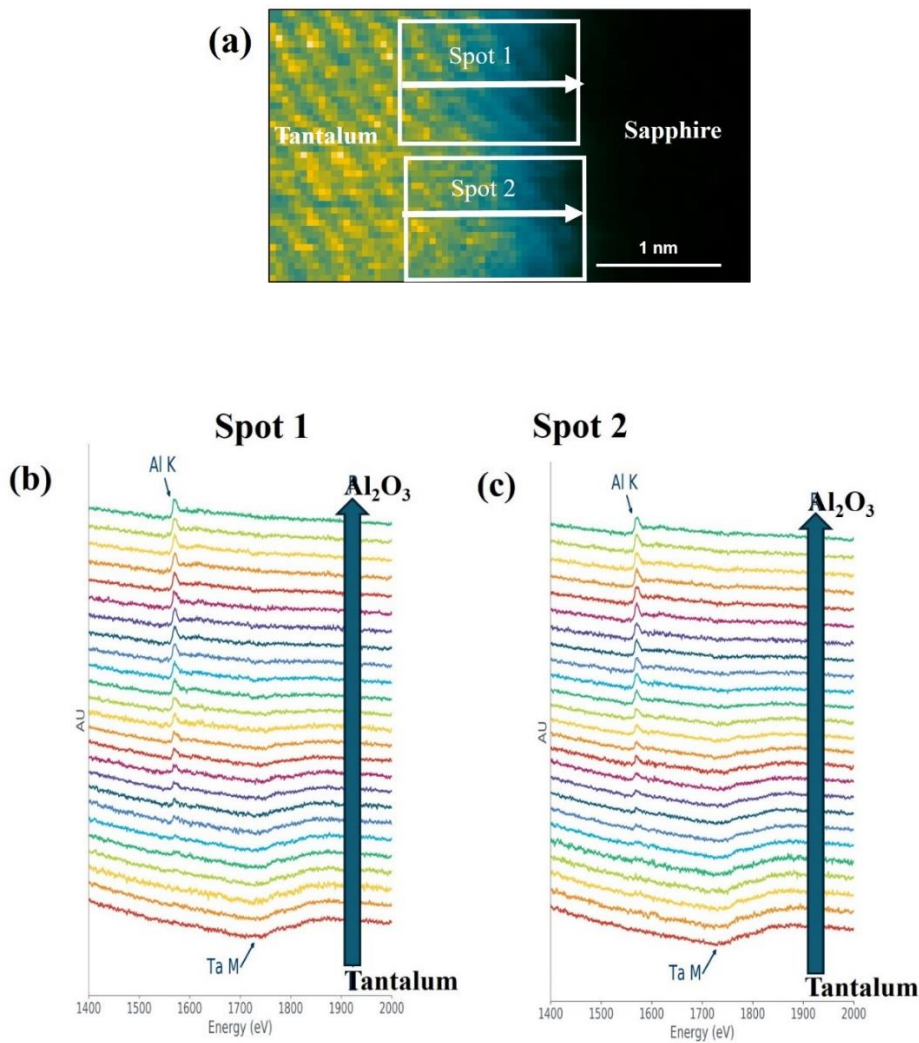

**Figure S3.** (a) HAADF-STEM false colored mass contrast imaging analysis of the Ta film on the substrate-metal interface and (b, c) the corresponding EELS scans of the highlighted regions Spot 1 and Spot 2. And the arrows denote the EELS scanning direction.

The VEXPS technique quantifies the chemical profile of the Ta film's surface region, correlating the surface oxide parameters with fitted XRR values. By varying the incident photon energy from low (690 eV) to high (4500 eV) (**Figure S4 a-f**), the kinetic energy of emitted photoelectrons can be varied, allowing depth measurements up to about 10 nm. **Figure S4 g-h** shows that below a thickness of 2-2.5 nm from the top, most of the Ta film remains in a pure metallic state.

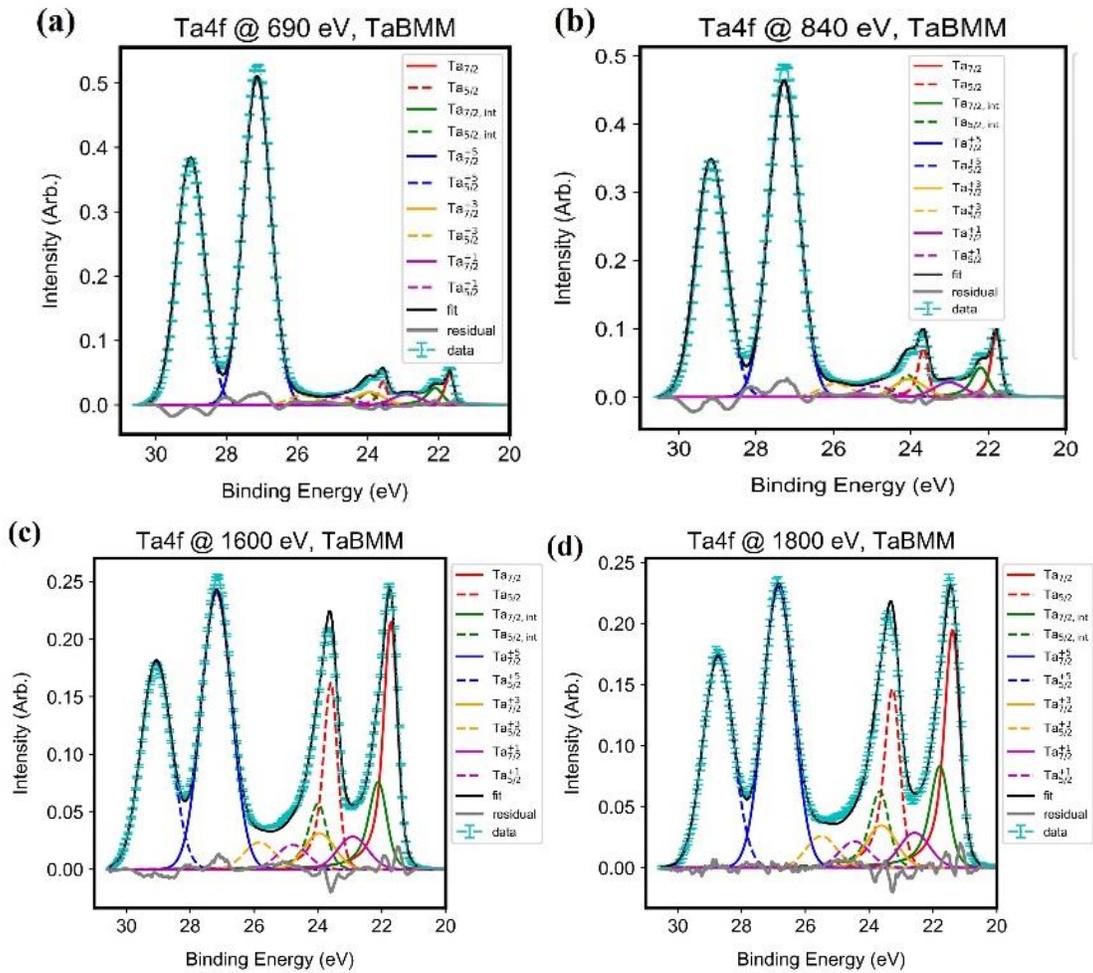

**Figure S4.** VEXPS measurements of Ta films at various incident photon energies.

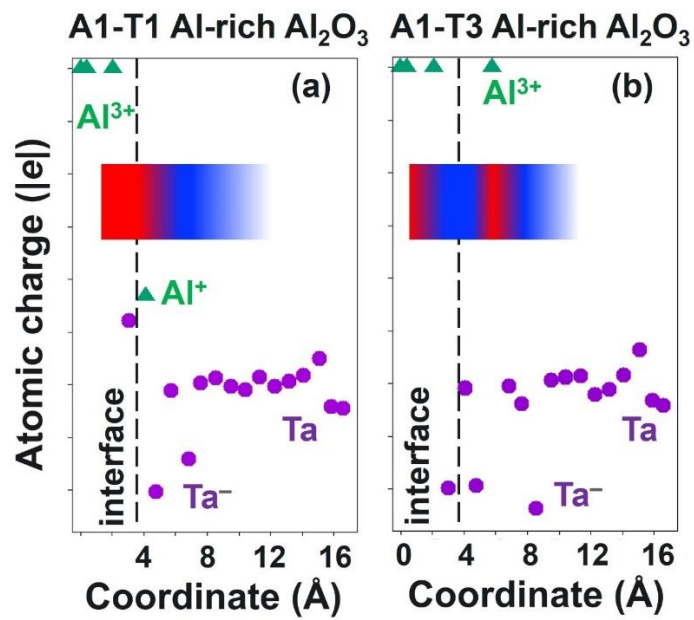

**Figure S5.** Charge distribution in selected Ta on Al-rich $Al_2O_3$ systems represented by layer-averaged Bader atomic charges of the Al and Ta species: intermixed configurations A1-T1 (a) and A1-T3 (b). The locations of Ta/$Al_2O_3$ interfaces are indicated with vertical dashed lines. The colored horizontal bars illustrate the variations of positive (red) and negative (blue) excess charge induced by the Al-Ta intermixing near the interface.

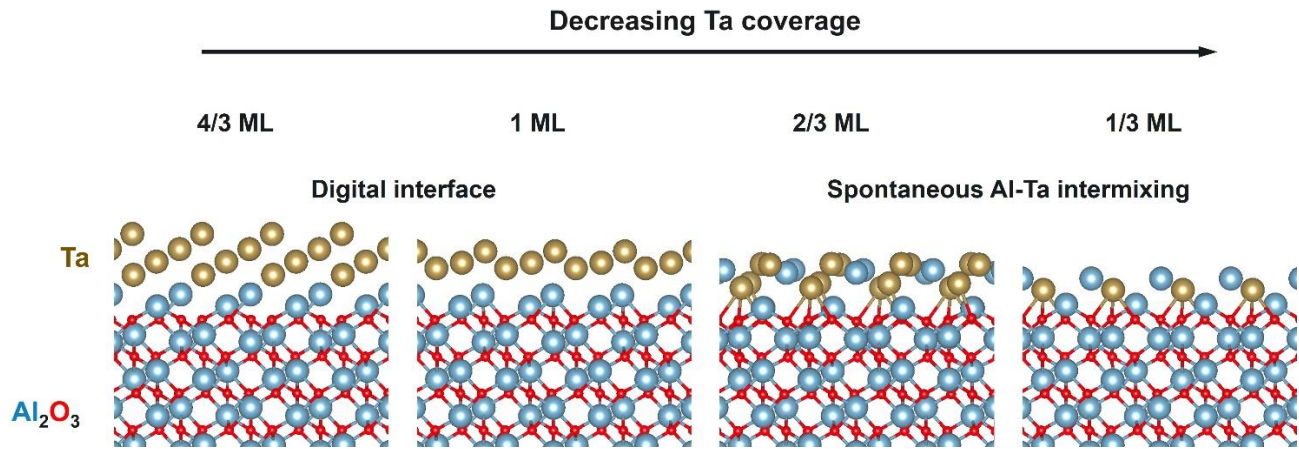

**Figure S6.** Layer-by-layer decreasing of the Ta coverage on the Al-rich surface causes a spontaneous Al-Ta rearrangement at the Ta coverage of ~2/3 ML. This observation indicates that in the reverse process, i.e., Ta deposition on the Al-rich $Al_2O_3$ (0001) surface, the surface Al and Ta atoms intermix at the early stages of Ta deposition resulting in the dissolution of the top al plane into the Ta film.

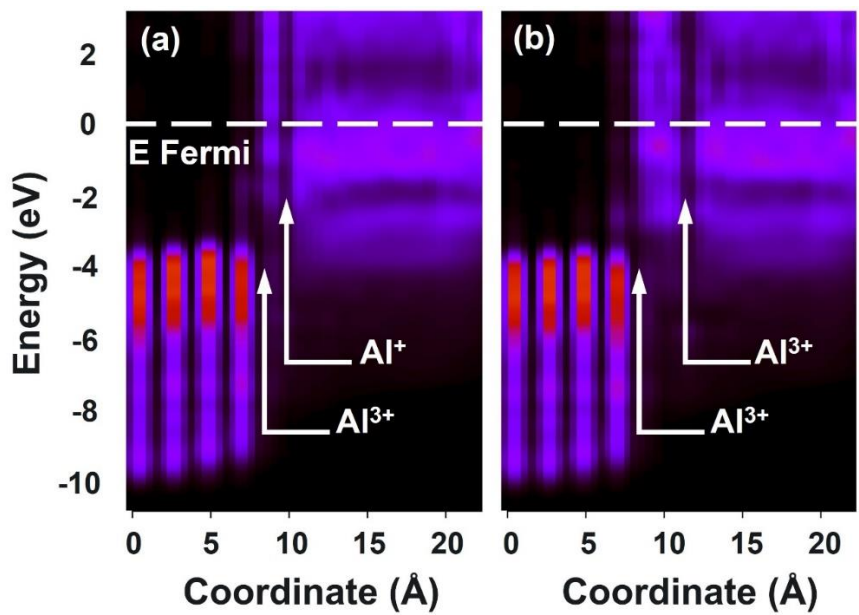

**Figure S7.** Heatmap representation of one-electron density of states calculated for selected Ta/Al$_2$O$_3$ systems and projected on the Al, O, and Ta atomic planes. Ta on Al-rich Al$_2$O$_3$: intermixed configurations A1-T1 (a) and A1-T3 (b), where A1 and T1 are the first Al and Ta planes near the interface, respectively, and T3 is the third Ta plane from the interface. All energy scales are aligned so that Fermi energy is at 0 eV. The substitutional Al impurities in Ta, arranged in planar configurations in (a) and (b), induce DOS depletion in the Ta film.

**Table S1. Comparison of findings on pure Ta (BCC structure) superconducting films.**

| Substrate | Deposition method | Film orientation (out of plane direction) | Ta thickness (nm) | Substrate treatment & deposition conditions | Tc [K] | RRR value |
|---|---|---|---|---|---|---|
| Sapphire (0001) [8] | DC magnetron sputtering | Ta (110) | 120 | Annealed up to 1100 °C (prior to deposition) | 4.2 | 4.5 |
| Sapphire (11$\bar{2}$0) [9] | RF magnetron sputtering | Ta (110) | 150 | Deposited at 750 °C | 4.18 | 7.5 |
| Sapphire (0001) [10] | RF magnetron sputtering | Ta (222) | 120 | Deposited at 750 °C | 4.3 | 19.36 |
| Sapphire (11$\bar{2}$0) [10] | RF magnetron sputtering | Ta (110) | 120 | Deposited at 750 °C | 4.28 | 7.48 |
| Sapphire (11$\bar{2}$0) [11] | Molecular beam epitaxy | Ta (110) | 30 | 200 °C for 2 h and followed by 850 °C for 0.5 h to remove surface | 4.12 | 9.53 |
| | | | | contamination (prior to deposition) and Ta deposited at 550 °C | | |
| Silicon (100) [12] | DC magnetron sputtering | Ta (110) | Unknown | cleaned inside the growth chamber at 500 °C (prior to deposition), then $TiN_x$ buffer layers and Ta deposited at 500 °C | 3.9 | 3.85 |
| Silicon (100) [13] | Sputtering | Ta (110) | 100 | Deposited at 400 °C | 2.9 | Unknown |
| Sapphire (0001) * | RF magnetron sputtering | Ta (222) | 30 | Deposited at 750 °C | 3.84 | 4.98 |

*This work